\documentclass[12pt,preprint]{aastex}

\usepackage{psfig}

	\newcommand{\Msun}{M$_{\odot}$} 
	\newcommand{\Lsun}{L$_{\odot}$}   
        \newcommand{\degrees}{$^{\circ}$~}

\shortauthors{Brooks et al.}
\shorttitle{A parsec-scale flow associated with IRAS 16547$-$4247}
\slugcomment{submitted to {\em The Astrophysical Journal Letters}}
\begin{document}


\title{A parsec-scale flow associated with the IRAS 16547$-$4247 radio
	jet}
\author{Kate J. Brooks}
\affil{Departamento de Astronom\'{\i}a, Universidad de Chile,
Casilla 36-D, Santiago, Chile}
\affil{and European Southern Observatory, Casilla 19001 Santiago, Chile}

\author{Guido Garay}
\affil{Departamento de Astronom\'{\i}a, Universidad de Chile,
Casilla 36-D, Santiago, Chile}

\author{Diego Mardones}
\affil{Departamento de Astronom\'{\i}a, Universidad de Chile,
Casilla 36-D, Santiago, Chile}

\and

\author{Leonardo Bronfman}
\affil{Departamento de Astronom\'{\i}a, Universidad de Chile,
Casilla 36-D, Santiago, Chile}

\begin{abstract}

IRAS 16547$-$4247 is the most luminous ($6.2 \times 10^4$~\Lsun) embedded
young stellar object known to harbor a thermal radio jet. We report the
discovery using VLT-ISAAC of a chain of H$_2$ 2.12 $\mu$m emission knots
that trace a collimated flow extending over 1.5 pc.  The alignment of the
H$_2$ flow and the central location of the radio jet implies that these
phenomena are intimately linked. We have also detected using TIMMI2 an
isolated, unresolved 12 $\mu$m infrared source towards the radio jet . Our
findings affirm that IRAS 16547$-$4247 is excited by a single O-type star
that is driving a collimated jet. We argue that the accretion mechanism
which produces jets in low-mass star formation also operates in the higher
mass regime.

\end{abstract}


\section{Introduction}

Whether massive stars (B and O-type) form via an accretion process similar
to that for low-mass stars (\citealt{Osorio99}; \citealt{McKee03}) or
instead via collisions with lower-mass stars \citep{Bonnell98} is currently
under debate.  Highly collimated radio jets (\citealt{Rodriguez97};
\citealt{Anglada98}) and Herbig-Haro (HH) flows \citep{Reipurth02} are
frequently observed towards young low-mass stars. It is now widely accepted
that jets are intimately linked to the accretion process and the formation
of both bipolar molecular outflows and HH flows \citep[see review
by][]{Reipurth01}. The role of collimated jets in massive stars (M $>$ 8
M$_{\odot}$) is less certain. In this higher mass regime observations of
jets are difficult, primarily because of the greater distances involved and
because the evolutionary time scales of such jets are expected to be much
shorter.

\citet{Garay03} recently reported the discovery of a triple radio continuum
source associated with IRAS 16547$-$4247, a young stellar object with a
bolometric luminosity of $6.2 \times 10^4$~\Lsun, equivalent to that of a
single O8 ZAMS star. The three radio components are aligned in a
southeast-northwest direction with the outer components (lobes)
symmetrically separated from the central source by an angular distance of
$\sim$20 arcsec. The triple system is centred on the position of the IRAS
source and is within a 1.2-mm dust continuum emission core whose properties
are similar to other massive star-forming cores
(e.g. \citealt{Garay02}). The spectral indices between 1.4 and 8.6 GHz are
0.49 for the central source and $-0.61$ and $-0.33$ for the two outer
components. This is consistent with the central source being a thermal jet
and the two outer components being non-thermal emission arising from the
working surfaces of the jet as it interacts with the surrounding ambient
medium. This is the first reported case of a radio jet associated with a
young O-type star.

We have performed a series of infrared observations towards IRAS 16547$-$4247
to confirm the existence of a collimated flow and to investigate the nature
of the powering source.

\section{Observations}
\subsection{Near-infrared Data}

Near-infrared images were obtained at the European Southern Observatory in
Paranal, Chile, using the ISAAC short wave camera \citep{Cuby00} mounted
on the ANTU telescope of the Very Large Telescope (VLT). The short wave
camera is equipped with a $1024 \times 1024$ pixel$^2$ Rockwell HAWAII
Hg:Cd:Te array with a pixel scale of 0\arcsec.148 pixel$^{-1}$, giving a
field of view of 2.5\arcmin $\times$ 2.5\arcmin.  The observations were
performed in service mode during the night of 2002 August 5 under
photometric conditions and with a seeing of 0.6 arcsec. Narrow-band
images centred on IRAS 16547$-$4247 were obtained using filters for the 
emission lines H$_2$1--0 S(1) 2.12 $\mu$m and Br$\gamma$ 2.16 $\mu$m,
together with the adjacent continuum at 2.09 $\mu$m and 2.19 $\mu$m. The
total on-source integration times were 22 min for the H$_2$ 2.12 $\mu$m and 
2.09 $\mu$m filters and 17 min for the Br$\gamma$ 2.16 $\mu$m and 2.19
$\mu$m filters.

Each frame was acquired using a jitter procedure with individual exposures
of $\approx$ 2 min per frame.  For each filter, the stack of individual
frames were first corrected for instrumental effects such as flat-field and
bias effects and then aligned using a cross-correlation method and
averaged. The sky contribution was determined from median-averaging the
stack of non-aligned frames and subsequently subtracting the
result. Astrometric calibration was done by identifying common stars
between the final images and an image from the online STScI Digitised Sky
Survey. We estimate the positional uncertainty to be less than 1 arcsec.

Photometric calibration of the H$_2$ 2.12 $\mu$m data was achieved by
observing HIP 080178. The final H$_2$ 2.12 $\mu$m image has a 1-$\sigma$
rms noise level of $1\times 10^{-16}$ erg s$^{-1}$ cm$^{-2}$
arcsec$^{-2}$. The 2.09 $\mu$m image was utilised to remove the continuum
contribution (stars) from the final H$_2$ 2.12 $\mu$m image. A perfect
subtraction is impossible because of small misalignments and differences in
the PSF between the two filters.

\subsection{Mid-infrared Data}

Mid-infrared images were obtained at the European Southern Observatory in
La Silla, Chile, using the TIMMI2 mid-infrared camera \citep{Reimann00}
mounted on the 3.6-m telescope. TIMMI2 is equipped with a $320 \times 240$
pixel$^2$ Raytheon Si:As array. A pixel scale of 0.3\arcsec pixel$^{-1}$
was used, giving a field of view of 96\arcsec $\times$ 72\arcsec. The
observations were made on the night of 2002 August 10 under photometric
conditions and with a seeing of 0.6\arcsec. Images were centred on IRAS
16547$-$4247 and obtained using the N11.9 $\mu$m filter. 

A chop-nod procedure was used with a nod duration of $\approx$ 4 min. The
chop and nod amplitudes were 15 arcsec, both oriented
north--south. Corrections for sky contribution and instrumental effects
were made by computing the double-difference of each chop-nod cycle. An
additional bias correction was applied to each of the chopped images. The
final image was constructed by shifting and adding all sources in each
chop-nod frame. The total on-source integration time was equivalent to 22
min.

Photometric calibration was achieved by observing HD 178345 and HD
4128. The final TIMMI2 image has a 1-$\sigma$ rms noise level of 0.02 Jy
arcsec$^{-2}$. No astrometric calibration was possible since there were no
other sources in the field. We estimate the telescope pointing accuracy to
be less than 5 arcsec.

\section{Results \& Discussion} 
\subsection{H$_2$ 2.12 $\mu$m emission}

\begin{figure}
    \centerline{\psfig{figure={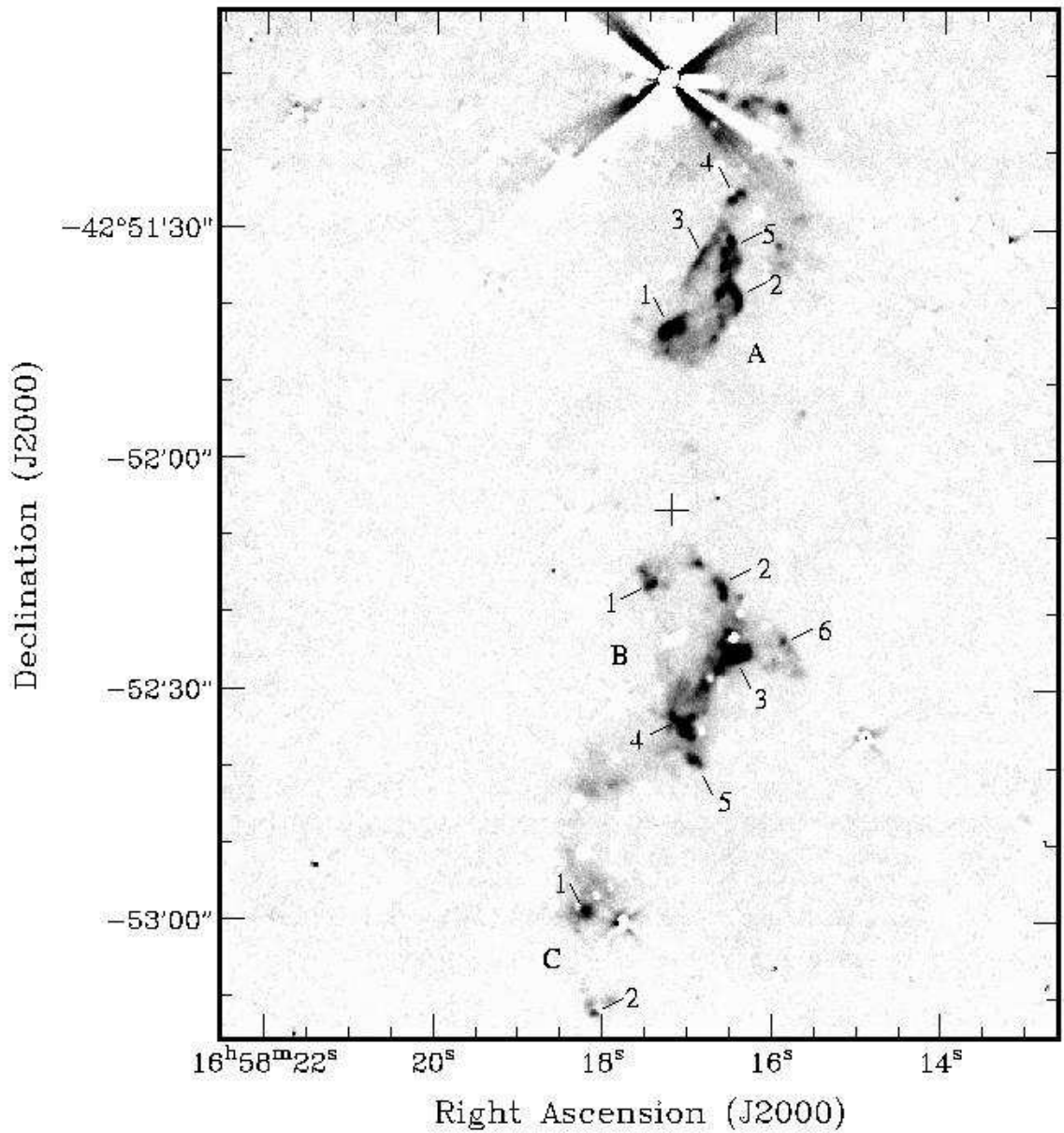},width=\textwidth}} 
    \caption{\label{plotone}ISAAC H$_2$ 2.12 $\mu$m emission image (continuum
    subtracted). A number of bright emission knots are labeled and their
    fluxes are listed in Table 1. The location of the thermal radio
    jet detected by Garay et al. (2003) is denoted by a cross (+). Artifacts at the top of the
    image are the result of an imperfect subtraction of the brightest
    star in the field. } 
    \end{figure}

Fig.~\ref{plotone} shows a map of the H$_2$ emission towards IRAS 16547$-$4247.
There is a complex chain of emission with three major concentrations
(labeled A to C). Several of the brightest emission knots within each
concentration have been labeled and their coordinates and fluxes are given
in Table~\ref{tableone}. The projected distance between the two outermost
knots (A4 and C2) is 110 arcsec (1.5 pc at the distance of 2.9 kpc,
Bronfman private communication). Both of these knots are
approximately symmetrically offset from the radio jet detected by
\citet{Garay03}. No emission arising from the Br$\gamma$ line was detected.

The H$_2$  emission has the morphological characteristics of HH
objects arising from the interaction of a collimated flow with the ambient
medium. Concentration A has several knots in the shape of bow-shocks, all
pointing away from the direction of the radio jet. Their arrangement is
consistent with an elongated outflow cavity. Another series of emission
knots may exist further north but is difficult to distinguish against the
artifacts from the bright star.  The morphology of concentration B is more
complex and consists of two main emission structures. Both structures
appear to delineate flows originating from a direction that is skewed from
the location of the radio jet: one in a northeast--southwest direction (B1,
B2, and B6) and one in a north--south direction (B3, B4 and B5).  This may
be evidence of additional outflows from less-massive stars or an indication
of precession of the flow originating from the detected radio jet. There
are fewer bright emission knots in concentration C and the morphology is
less well-defined. There are a couple of faint filaments in the shape of
bow-shocks pointing away from the direction of the radio jet. These are
most likely part of the counter-flow to concentration A.

\begin{figure}
    \centerline{\psfig{figure={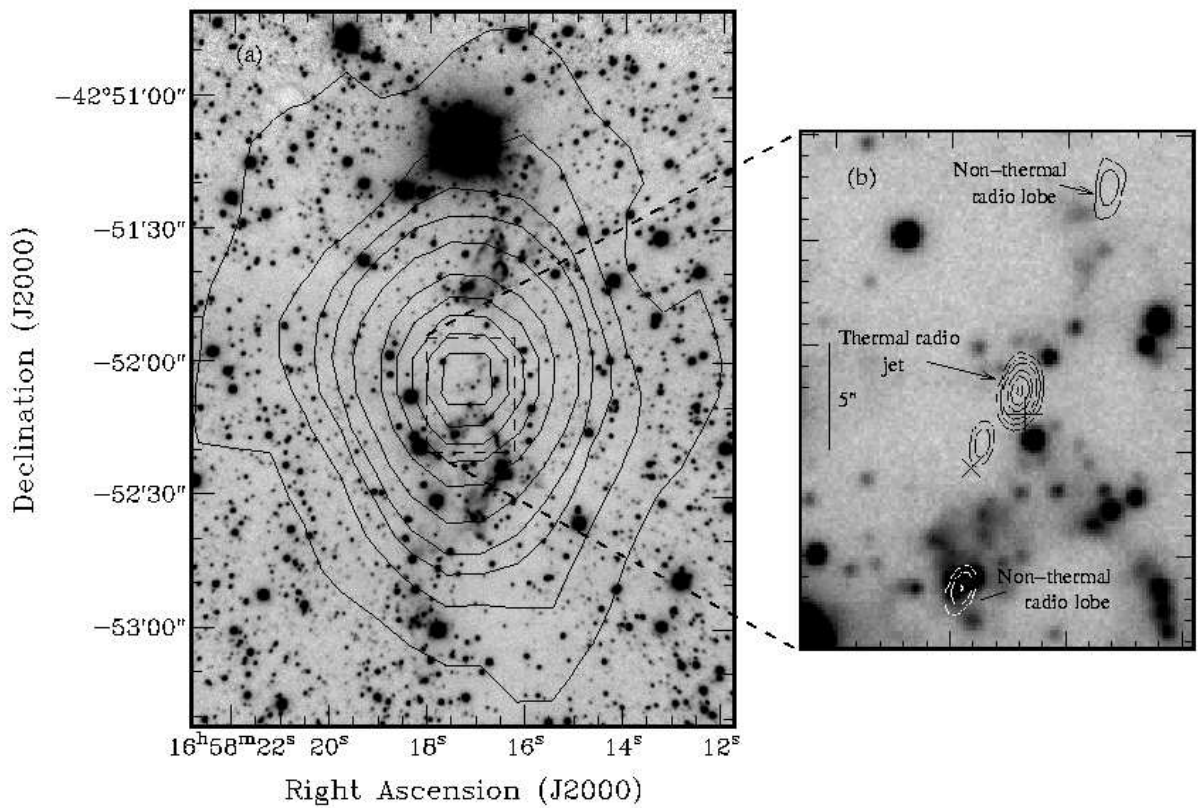},width=\textwidth}} 
    \caption{\label{plottwo} ISAAC H$_2$ 2.12 $\mu$m emission image (no continuum
    subtraction) shown in grey scale. (a) Overlaid with 1.2-mm dust
    continuum emission from Garay et al. (2003). Contour levels are 0.25
    (5$\sigma$), 0.5, 0.75, 1, 1.5, 2.5, 3.5, 4.5, 5.5, 6.5 Jy beam$^{-1}$.
    (b) Overlaid with 8.6-GHz continuum emission from Garay et
    al. 2003. Contour levels are 0.35 (5$\sigma$), 0.6, 1.1, 2.3, 3.5, 5
    mJy beam$^{-1}$. Also shown are the OH maser position (+) from
    Caswell (1998) and the H$_2$O maser position ($\times$) from
    Forster \& Caswell (1989). There is a positional uncertainty of 1\arcsec\ in
    registering the H$_2$ 2.12 $\mu$m and 8.6-GHz continuum data.} 
    \end{figure}

Fig.~\ref{plottwo} illustrates the comparison between the H$_2$ emission
(without any continuum subtraction) and (a) the 1.2-mm dust continuum
emission and (b) the 8.6-GHz continuum emission taken from
\citet{Garay03}. The actual thermal radio jet corresponds to the brightest
8.6-GHz emission component. There is a fainter source offset to the
southeast and whose spectral index is not known. It is not certain what
role (if any) this source plays. The chain of H$_2$ emission is oriented in
the same direction as the triple radio source and is contained within the
molecular core traced by the 1.2-mm continuum emission. One of the H$_2$
emission knots (B1) is associated with the southern non-thermal radio
component. The striking alignment of the collimated flow delineated by the
H$_2$ emission and the central location of the radio jet implies that these
phenomena are coupled. It is reasonable to assume that the radio jet is the
driving force of the collimated flow. No stellar counterpart to the radio
jet was detected in the narrow-band near-infrared images (e.g. see
Fig.~\ref{plottwo}b).

IRAS 16547$-$4247 is reminiscent of another young massive stellar object,
HH80-81 ($\approx 1.7 \times 10^4$~\Lsun\ at 1.7 kpc). HH80-81 contains a
thermal radio jet centred on a collimated HH flow that extends over 5 pc
\citep{Marti93}. Evidence for non-thermal radio emission has also been
reported towards one of the HH emission knots.  Up until now this was the
most luminous young stellar object known to have a collimated jet.

\subsection{11.9-$\mu$m emission}

The 11.9-$\mu$m image obtained in this study reveals a single unresolved
($<$ 0.6 \arcsec) emission source with a flux of 0.28 Jy, albeit there may
be less-luminous objects present that are below the sensitivity limit of
the observations. Within the positional uncertainty of the mid-infrared
data, this source is coincident with the radio jet. The flux is consistent
with the spectral energy distribution for IRAS 16547$-$4247 shown in
\citet{Garay03}.

According to the relationship between jet radio-luminosity and outflow
momentum rate reported by \citet{Anglada98}, extrapolated to the high-mass
regime, the IRAS 16547--4247 outflow should have a momentum rate of
0.2~\Msun\ yr$^{-1}$ km~s$^{-1}$. This mechanical force can only be
supplied by a driving source with a luminosity of $ \approx 10^5$ \Lsun\
\citep{Beuther02}. The measured luminosity for IRAS 16547--4247 is $6.2
\times 10^4$~\Lsun. Therefore we argue that the radio jet and the H$_2$
collimated flow are powered by the same single luminous
object. Furthermore, it is reasonable to suppose that the detected
11.9-$\mu$m emission source is associated with this luminous object. 

\section{Conclusions} 

IRAS 16547$-$4247 is the most luminous embedded pre-main sequence source in
the Galaxy known to harbor a thermal radio jet. We have detected a chain of
H$_2$ 2.12 $\mu$m emission knots towards IRAS 16547$-$4247 that delineate a
collimated flow extending over 1.5 pc. The geometry of the flow implies
that it is driven by the thermal jet. We have also identified an isolated
unresolved mid-infrared object associated with the jet that is likely to be
responsible for the excitation of IRAS 16547$-$4247. If the luminosity of
IRAS 16547--4247 comes from a single object it would have the spectral type
O8. The finding of a jet and collimated flow towards such a massive object
supports the accretion scenario for the formation of stars across the entire
mass spectrum. 

\acknowledgements 

We thank Vanessa Doublier, Rachel Johnson, Nathan Smith and Michael Sterzik
for their help with observations and data reduction. The ISAAC data were
obtained through the ESO Director's Discretionary Time Program. This work
has been partly funded by the Chilean Centro de Astrof\'{\i}sica FONDAP N$^o$15010003.


\clearpage

\newpage
\clearpage

\begin{deluxetable}{ccccc} 
\tablecolumns{5} 
\tablewidth{0pc} 
\tablecaption{\label{tableone}Flux parameters for the H$_2$ 2.12 $\mu$m
knots identified in Fig.~1.} 
\tablehead{ 
\colhead{}   &\multicolumn{2}{c}{Peak Position} &  Peak Flux    & Total
Flux\tablenotemark{a}\\
\colhead{}   & RA(J2000)  & Dec(J2000) & $\times 10^{-15}$ erg s$^{-1}$ 
        & $\times 10^{-14}$ \\
\colhead{}   & $\rm ^{h}~^{m}~^{s}$ & \degrees  \arcmin ~ \arcsec
        & cm$^{-2}$ arcsec$^{-2}$ & erg s$^{-1}$ cm$^{-2}$
}
\tablenotetext{a}{Each integrated over $\approx$ 4\arcsec $\times$ 4\arcsec\ box.}
\startdata 

	A1
	& 16 58 17.2
	& $-$42 51 43
	& 11
	& 5.0
	\\

	A2
	& 16 58 16.5 
	& $-$42 51 38
	& 5.1
	& 4.5
	\\

	A3
	& 16 58 16.9
	& $-$42 51 34
	& 2.6
	& 1.4
	\\

	A4
	& 16 58 16.5
	& $-$42 51 26
	& 4.1
	& 1.3
\\	
	A5
	& 16 58 16.5
	& $-$42 51 31
	& 5.8
	& 3.6
	
\\

	B1
	& 16 58 17.4
	& $-$42 52 16
	& 3.3
	& 1.7
	\\
	
	B2
	& 16 58 16.6
	& $-$42 52 17
	& 3.6
	& 1.6
	\\
	
	B3
	& 16 58 16.4
	& $-$42 52 25
	& 31
	& 9.3
	\\

	B4
	& 16 58 17.0 
	& $-$42 52 35
	& 4.3
	& 3.7
	\\

	B5
	& 16 58 16.9
	& $-$42 52 39
	& 3.6
	& 1.1
	\\

	B6
	& 16 58 15.9
	& $-$42 52 24
	& 2.1
	& 0.4
\\

	C1
	& 16 58 18.2
	& $-$42 52 59
	& 3.4
	& 0.8
	\\
	
	C2
	& 16 58 18.1
	& $-$42 53 12
	& 2.1
	& 0.5
	\\
\enddata 
\end{deluxetable} 


\clearpage

\end{document}